\documentclass[journal,twocolumn,10pt]{IEEEtran}

\usepackage{amsmath,amssymb,amsthm}
\usepackage{graphicx}
\usepackage{algorithm}
\usepackage{algorithmicx}
\usepackage{algpseudocode}
\usepackage{xcolor}
\usepackage{tabularx}
\usepackage{threeparttable}

\newtheorem{lem}{Rule}
\usepackage{array}
\newcolumntype{L}[1]{>{\raggedright\let\newline\\\arraybackslash\hspace{0pt}}m{#1}}
\newcolumntype{C}[1]{>{\centering\let\newline\\\arraybackslash\hspace{0pt}}m{#1}}
\newcolumntype{R}[1]{>{\raggedleft\let\newline\\\arraybackslash\hspace{0pt}}m{#1}}

\usepackage[tight,footnotesize]{subfigure}
\PassOptionsToPackage{draft}{hyperref}
\ifCLASSINFOpdf
\else
\fi
\begin{document}

\title{On Coordinating Ultra-Dense Wireless Access Networks: Optimization Modeling, Algorithms and Insights}


\author{\IEEEauthorblockN{Antonis G. Gotsis and Angeliki Alexiou}\\
\IEEEauthorblockA{Department of Digital Systems, University of Piraeus, Greece\\
Email: \{agotsis,alexiou\}@unipi.gr}
}

\maketitle

\begin{abstract}
Network densification along with universal resources reuse is expected to play a key role in the realization of 5G radio access as an enabler for delivering most of the anticipated network capacity improvements. On the one hand, neither the expected additional spectrum allocation nor the forthcoming novel air-interface processing techniques will be sufficient for sustaining the anticipated exponentially-increasing mobile data traffic. On the other hand, enhanced ultra-dense infrastructure deployments are expected to provide remarkable capacity gains, regardless of the evolutionary or revolutionary approach followed towards 5G development. In this work, we thoroughly examine global network coordination as the main enabler for future 5G large dense small-cell deployments. We propose a powerful radio resources coordination framework through which interference management is handled network-wise and jointly over multiple dimensions. In particular, we explore strategies for pairing serving and served access nodes, partitioning the available network resources, as well as dynamically allocating power per pair, towards optimizing system performance and guaranteeing individual minimum performance levels. We develop new optimization formulations, providing network scaling performance upper bounds, along with lower complexity algorithmic solutions tailored to large networks. We apply the proposed solutions to dense network deployments, in order to obtain useful insights on network performance and optimization, such as rate scaling, infrastructure density, optimal bandwidth partitioning and spatial reuse factor optimization.
\end{abstract}



%
\IEEEpeerreviewmaketitle


\section{Introduction}\label{sec:Intro}
\subsection{Motivation -- Ultra Dense Networks}\label{sec:Intro_Motivation}
The mobile broadband era has arrived as the global deployment status of beyond 3G (HSPA+) and 4G (LTE/LTE-A) radio technologies suggests. With more than 900 HSPA and 200 LTE networks in service and 240 new networks in trial/planned phase all over the world~\cite{4GAmerNetdeployment}, pervasive and high-speed wireless access to Internet data services is now the case. At the same time market analyses forecast a 13-fold increase for global mobile data traffic by 2017~\cite{Cisco13} and more than 50x by 2020, posing serious challenges to current wireless access networks. To cope with this so-called ``capacity crunch", wireless technology stakeholders have already started to shape the evolution of the future radio access network of 2020 and beyond, also referred to as 5G~\cite{Ericsson5GWP13}.

From a wireless communications design perspective, the currently planned spectrum refarming or introduction of new carriers (e.g. in 3.5 GHz or even in the mmWave band) may not be able to cater for such traffic demand. Neither the anticipated improvement of spectral efficiency levels through a new air-interface and evolutionary physical-layer processing (e.g. Massive/3D MIMO, non-orthogonal multiple-access) can boost network capacity by more than two or three orders of magnitude compared to today's standards~\cite{AnSi13}. On the other hand, a heavy increase in infrastructure density along with universal resources reuse seems to be a promising solution for the capacity crunch~\cite{HwSo13}, as the access network is brought closer to the user and the radio resources are readily available practically ``everywhere".  Recently, this ultra-dense network deployment (UDN) trend is gaining ground as it promises significant capacity improvements, and is considered for inclusion in the upcoming LTE releases (12 and beyond)~\cite{AsDa13}.

Although network densification and universal resources reuse is a typical capacity increase strategy in the cellular paradigm, new challenges and issues arise when incorporated in the 5G framework. Heavy irregular infrastructure deployment of low-powered access nodes (ANs) leads to \emph{random topology networks}, for which interference conditions characterization becomes harsh. The combination of random topology infrastructure with highly inhomogeneous user terminals and devices (UEs) distributions, renders coverage and capacity prediction practically impossible. To this end, sophisticated across the network coordination of transmissions and resources allocation, realized by a centralized cloud-based architecture~\cite{NiCo13} seems to be the key enabler of UDN capacity scaling.

\subsection{Related Work}\label{sec:Intro_RelatedWork}
Recent views on the ultra dense small cell networking evolution are presented in~\cite{HwSo13} along with preliminary throughput per unit area performance results for ever-densified deployments. There are two limitations of this work: first, the adopted performance metric is the average data rate, which does not reflect possible QoS discrepancies across the network, and second, evaluations are based on a baseline ``ad-hoc-like" operation scheme, neglecting the potential benefits of intelligent coordination. From a resources management point of view, joint and network-wise handling of transmission parameters and medium access control (e.g. power levels, selection of serving access nodes, orthogonalization of different AN-UE paired transmissions) becomes highly critical in large UDNs. With ever-increasing UEs population over a certain area, and the ensued proportionate increase in infrastructure density, providing capacity scaling for increasing network sizes (defined as the number ANs and UEs over a serving area) becomes challenging. Such complex problems are usually looked at from an optimization perspective.

Starting from the sub-problem of \textit{optimizing power control} over networks forcing universal spatial reuse (that is, all transmissions occur simultaneously and over the same shared medium, without any orthogonalization), a rich amount of solutions can be found in the literature, covering optimality and implementation aspects. The problem is in general non-convex due to the interference terms coupling \cite{ChTa07}, hence either reformulation or approximation/relaxation approaches are followed. For example, in \cite{ChTa07} an approximated algorithm for maximizing the network sum-rate, based on successively solving a series of geometric (convex) problems is proposed, while in~\cite{KhTu12} the differential of convex functions (D.C.) programming framework is applied to the sum-rate and max-min rate problems. In \cite{GoAl13} and \cite{SuHo12} the optimal network-wise power allocation for the max-min problem is acquired with polynomial complexity through solving a series of linear programs (LPs). Finally in \cite{TaCh13}, a simple distributed online power control algorithm converging to optimality is proposed.

When \textit{power coordination} is \textit{jointly considered} with the per UE best serving AN selection (the known cell-selection problem, referred as AN-UE \textit{pairing} hereafter) the problem becomes significantly more challenging due to the combinatorial nature of UE-to-AN association decisions. Regarding the max-min rate optimization problem, the work in \cite{GoAl13} reformulates the original Mixed-Integer Non-Linear Program (MINLP) and provides a global optimal solution by solving a series (typical 10-20) of Integer Linear Programs (ILP). Similarly in \cite{SuHo12}, the optimal solution is provided utilizing concepts from combinatorial optimization, namely the hungarian algorithm, Moreover in \cite{QiZh13}, the system revenue maximization problem considering individual rate guarantees is tackled through the ILP optimization framework and a suboptimal algorithm utilizing Benders decomposition principles is also proposed.

So far full reuse of radio resources is considered. As demonstrated in \cite{GoAl13}, even for the optimal resources management solution, per-UE capacity is severely degraded with increasing network sizes, if common rate QoS (max-min fairness) is required across the network. To this end, \textit{partitioning}, that is fully or partially orthogonalizing the various AN-UE paired transmissions, becomes relevant. Partitioning can be employed in one or more domains, such as time, frequency/bandwidth, space, code, etc. (without loss of generality we will consider bandwidth partitioning hereafter). It controls interference by decreasing the spatial reuse of resources, thus increasing the achievable SINR levels at the expense of reduced bandwidth utilization.

This fundamental trade-off has been thoroughly examined from an analytical perspective in \cite{JiAn08} for an ad-hoc network, leveraging powerful stochastic geometry tools. Due to the analytical nature of this work several idealized/over-simplifying assumptions and non-adaptive communication strategies were employed (predetermined communication pairs, no power control, random partition selection, fixed common distance among transmitters-receivers etc.), not tailored to infrastructure-aided networks. Along the same lines, the authors in \cite{LiWu13} utilized stochastic geometry for obtaining performance and design insights for the average performance levels of future UDNs. An alternative path is to borrow ideas from multi-cell OFDMA resources management problems, due to its resemblance with the UDN coordination problem. For example, in \cite{VePr09} various optimization approaches for the weighted sum-rate maximization problem are proposed, while in \cite{Ne09,ChTa09} graph-based theory and practices are applied to the UEs partitioning problem. Nevertheless, neither of those is applicable to the system model assumed in this paper, since joint consideration of all problem dimensions is not taken into account.

\subsection{Contribution -- Paper Organization}\label{sec:Intro_Contribution}
It has become clear that the joint radio resources management and coordination problem is one of the key 5G technical challenges and only partially addressed in the literature so far. This problem needs to be tackled especially in the UDN context, in order to understand the capabilities and limitations of large dense infrastructure deployments. To this end the current work aims at exploring the key pillars for efficiently coordinating an ever-densified wireless access network, namely AN-to-UE pairing, fully or partially orthogonal partitioning of AN-UE paired transmissions, and dynamic power control per AN-UE pair. Thus, we develop a comprehensive mathematical programming modeling framework entailing various optimization choices, along with an algorithmic solutions ``toolbox". More specifically:
\begin{itemize}
  \item We formulate the joint pairing, partitioning and power (PPP) coordination problem which satisfies minimum service rate requirements or searches for the maximum common provided levels, and to the best of our knowledge we derive for the first time an exact solution for it. The Joint-PPP solution provides a performance benchmark for any related network-wise coordination algorithm but suffers from complexity scaling issues. The special case of joint partitioning and power coordination given known or a-priori forced pairing (e.g. each UE served by its closest AN) is also addressed due to its wide applicability. Both exact optimization formulations leverage ILP, a mature optimization area, for which powerful solvers are available.
  \item We develop an algorithmic framework providing suboptimal solutions with reasonable complexity, appropriate for large dense deployments coordination comprising tens or even hundreds of infrastructure and served nodes. The framework is based on decomposing the Joint-PPP problem into multiple problem components, utilizing simple yet effective greedy-type allocation rules for the combinatorial (pairing, partitioning) decisions, as well as employing closed-form expressions for continuous variables (power coordination) optimization. Exploiting this framework, we introduce two suboptimal-PPP algorithms trading-off performance and complexity, depending on the joint or disjoint consideration of partitioning and power control.
  \item We apply the optimal and suboptimal solutions to several UDN system setups, showcasing various degrees of performance (optimal vs suboptimal), computational complexity (polynomial vs exponential) and implementation capabilities (fully vs partially centralized). We also exploit the algorithms to obtain insights for future UDN parametrization and design, namely the dependence of partitioning on infrastructure density, optimal partitioning size tuning, and the achieved performance scaling vs network densification.
\end{itemize}
The rest of the paper is structured as follows. In Section~\ref{sec:SystemModel} notation, assumptions, and system model are stated, followed by the Joint-PPP problem definition. In Section~\ref{sec:OPTIMAL_JOINTPPP} we describe the reformulation of the Joint-PPP problem to an optimization model that allows us to obtain the global optimal coordination decisions. As a special case, the joint partitioning and power coordination problem for fixed AN-UE pairing is also considered. Next, in Section~\ref{sec:SuboptimalPPP} we describe the Suboptimal-PPP framework along with two algorithmic versions. In Section~\ref{sec:Results} we validate the proposed solutions as well as discuss performance and parametrization issues for UDNs through simulation experiments. Finally, Section~\ref{sec:Conclusion} concludes our work and states possible future research directions.
\section{System Model and Problem Definition}\label{sec:SystemModel}

\subsection{Notation and Assumptions}\label{sec:SystemModel_a}
We consider a randomly deployed wireless network comprising a set $\mathcal{M}$ of single-antenna ANs (typical infrastructure elements such as small-cells or access points), serving a set $\mathcal{K}$ of randomly dropped single-antenna terminals/devices, i.e. user equipment (UEs), residing in a typical macro-cell area. Let $M$ and $K$ the cardinality of the above sets. Differently from the classic cellular paradigm, in the explored UDN context, we consider at least as many ANs as UEs, i.e. $M \geq K$, corresponding to a densified small-cell use-case, catering for full bandwidth reuse, as each UE is associated with its ``own" corresponding AN. Such a use-case is considered in the standardization of upcoming 3GPP releases~\cite{NaNa13} and in 5G research projects~\cite{METISD11}. We also regard a set $\mathcal{N}$ (with cardinality $N$) of mutually orthogonal network partitions formed by some orthogonalization approach (e.g. FDMA, TDMA, etc.). Without loss of generality, we will assume FDMA orthogonalization hereafter. In practical terms, each partition may correspond to a single LTE physical resource block (PRB), a group of PRBs, or a carrier (in the context of carrier aggregation and cross-carrier scheduling/enhanced inter-cell interference coordination in LTE-A). For $N=1$, universal frequency reuse for all AN-UE pairs is forced (no-partitioning or ``full-spatial-reuse" baseline case), whereas for $N=K$, no inter-pair interference occurs by having each partition associated with a unique UE (``full-orthogonalization" baseline case). All other cases ($1 < N < K$) are classified as partially partitioned.

In addition, $g_{km}$ denotes the noise-power normalized channel gain of the $k^{\text{th}}$ UE to the $m^{\text{th}}$ AN. We further introduce a binary (0-1) indicator variable $\rho_{kmn}, \forall k,m,n$, which is equal to 1 if the $k^{\text{th}}$ UE is served by the $m^{\text{th}}$ AN at the $n^{\text{th}}$ partition. Accordingly, $p_{kn}$ denotes the power devoted to the $k^{\text{th}}$-UE transmission, that is the power allocated to the AN serving the UE, at the $n^{\text{th}}$ partition. We also consider a Shannon-based PHY abstraction~\cite[10.4.1]{HoTo11} for translating achieved SINR levels to spectral-efficiency or rate. Then, assuming gaussian signaling and treating interference as noise, the achieved SINR for the arbitrary $k^{\text{th}}$ UE over the $n^{\text{th}}$ partition ($\gamma _{kn}$) and the achieved rate per UE ($R_k$) are given as follows:
\begin{equation}\label{eq:BasicSINRexpr}
\begin{gathered}
  {\gamma _{kn}} = \frac{p_{kn} \cdot {\sum\limits_{m \in \mathcal{M}} {{g_{km}} \cdot {\rho _{kmn}} } }}{{1 + p_{in} \cdot \sum\limits_{i \in \mathcal{K}\backslash \left\{ k \right\}} {\sum\limits_{m \in \mathcal{M}} {{g_{km}} \cdot {\rho _{imn}} } } }}, \hfill \\
  {R_k}\left( {{\text{bps/Hz}}} \right) = \left( {{1 \mathord{\left/
 {\vphantom {1 N}} \right.
 \kern-\nulldelimiterspace} N}} \right) \cdot {\log _2}\gamma _k^{eff}, \gamma _k^{eff} = \prod\limits_{n \in \mathcal{N}} {\left( {1 + {\gamma _{kn}}} \right)} , \hfill \\
 \end{gathered}
\end{equation}
where $\gamma _k^{eff}$ denotes the effective SINR, namely the log-inverted average SINR over all partitions.

\subsection{Problem Definition and Formulation}\label{sec:SystemModel_b}
\begin{figure}
\centering
\subfigure[No Partitioning - Full Spatial Reuse of Bandwidth]{%
\includegraphics[scale=.38]{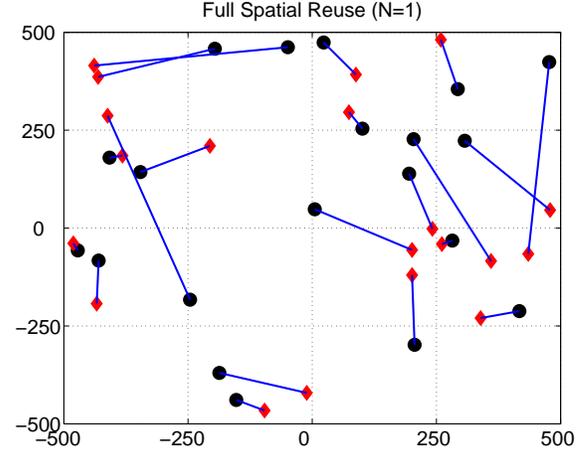}}
\label{fig:a}
\subfigure[Partitioned Reuse of Bandwidth (2 partitions)]{%
\includegraphics[scale=.38]{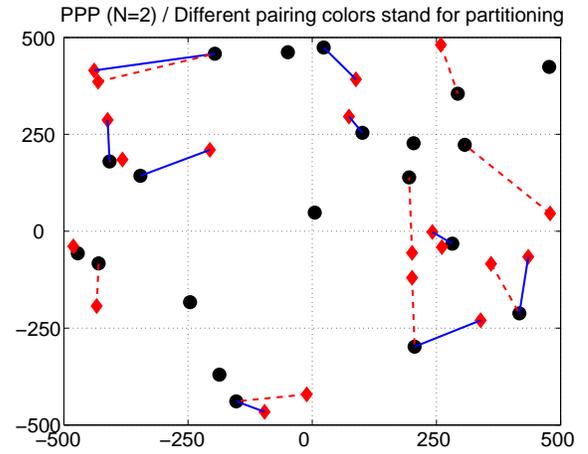}}
\label{fig:b}
\subfigure[PPP based on ``Cellular-Paradigm" Pairing. Note that the geographic limits of ANs or ``voronoi cells" are meaningful in this scenario and shown.]{%
\includegraphics[scale=.38]{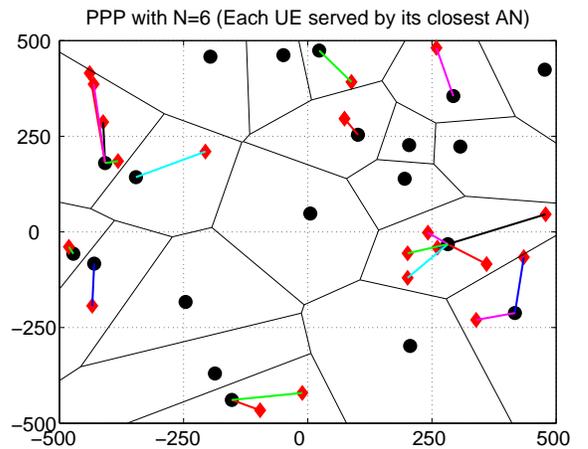}}
\label{fig:c}
\caption{A random densified network deployment and potential PPP solutions (circles stand for ANs, diamonds for UEs, and connecting lines for AN-UE pairing)}
\label{fig:Deployment}
\end{figure}

From a QoS perspective, a common attainable rate per UE across the whole network is targeted. This is inspired by the latest LTE-A target requirements~\cite{HoTo11}, where cell-edge UEs experiencing deteriorated radio quality conditions should enjoy high bit-rate services, comparable to those of served UEs that lie around the cell centre. This inspires a ``max-min" optimization which turns out to provide a common rate. Although not very flexible, such a QoS criterion provides also the maximum (common or individual) feasible service levels in a dense network deployment, something that has been overlooked in previous studies (e.g. in \cite{HwSo13} where average performance levels, with potentially large QoS discrepancies, were provided).  Towards this purpose, we search for optimal and efficient strategies for reusing (under a close-to-unity factor) frequency resources, aiming at maximizing network spectral efficiency but at the same keeping the experienced interference at tolerable levels. In the examined UDN context, a network coordinator will be responsible for:
\begin{itemize}
  \item \textbf{Pairing}, i.e. the association of each UE with a serving AN in each partition. Note that since cooperation in the data-plane (that is Network-MIMO or CoMP) is not considered, in each partition each AN serves exactly one UE or is turned-off when no UE is associated with it.
  \item \textbf{Partitioning}, i.e. the selection and distribution of UEs subsets to the orthogonal partitions. Due to complexity reduction reasons a UE is assumed to be assigned to a single partition.  Multiple AN-UE pairs are active in each partition and each AN could serve different UEs in different partitions.
  \item \textbf{Power Coordination}, the control of the transmission power for each AN-UE pair, which impacts both useful and interfering signals received power levels.
\end{itemize}
Fig.~\ref{fig:Deployment} illustrates a typical $1\text{ km}^2$ random dense network deployment comprising $20$ ANs and $20$ UEs, and three possible PPP solutions, that is, full-spatial-reuse (corresponding to maximum bandwidth utilization at the expense of decreasing SINR), partitioning with $N=2$, and a special case (``Cellular-Paradigm"), where each UE is forced to be served by its closest UE and AN-UE pairs with the same AN-end are orthogonalized (leading to lower bandwidth utilization).
%

We now cast network-wise coordination as an optimization problem. We consider the three-dimensional binary matrix {\boldmath{$\rho$}} $\in {\left\{ {0,1} \right\}^{K \times M \times N}}$ capturing both pairing and partitioning decisions, and the power allocation matrix ${\mathbf{p}} \in {\mathbb{R}}^{K \times N}_{+}$ capturing the power levels, each one upper bounded by $p_{max}$ reflecting per-UE or equivalently per-AN power constraints under the pairing/partitioning restrictions. Then maximizing the common attainable SINR per UE for a given bandwidth partitioning size $N$, corresponds to solving the optimization problem (P1):

\begin{subequations}\label{eq:PForig}
\small
\begin{gather}
  {\text{P1}}:{\theta ^*} = \arg \mathop {\max \theta }\limits_{\text{\boldmath{$\rho$}},{\mathbf{p}}}  \hfill \nonumber\\
  \text{s.t.  } \gamma _k^{eff} \geqslant \theta ,\forall k, \hfill \label{eq:PForig_MinSINR}\\
  \sum\limits_{m \in \mathcal{M}} {{\rho _{kmn}}}  \leqslant 1,\forall k,\forall n,\sum\limits_{k \in \mathcal{K}} {{\rho _{kmn}}}  \leqslant 1,\forall m,\forall n, \hfill \nonumber \\ \sum\limits_{m \in \mathcal{M}} {\sum\limits_{n \in \mathcal{N}} {{\rho _{kmn}}} }  = 1,\forall k, \hfill \label{eq:PForig_Connectivity}\\
  0 \leqslant {p_{kn}} \leqslant {p_{\max }},\forall k,{\rho _{kmn}} \in \left\{ {0,1} \right\},\forall k,\forall m,\forall n \label{eq:PForig_VarSpace}.
\end{gather}
\end{subequations}
The optimum supported common rate $R_{common}^*$ is then simply given by $R_{common}^* = \left( {{1 \mathord{\left/
 {\vphantom {1 N}} \right.
 \kern-\nulldelimiterspace} N}} \right) \cdot {\log _2}{\theta ^*}$. Constraints \eqref{eq:PForig_MinSINR} guarantee that no better effective SINR for one UE exists, which at the same time does not deteriorate the SINR of another UE. Constraints \eqref{eq:PForig_Connectivity} provide the feasible connectivity/pairing requirements, that is, each UE is served by at most one AN, each AN is serving at most one UE, as well as each UE is assigned to a single partition, respectively. Finally, constraints \eqref{eq:PForig_VarSpace} define the design variables feasible space. Note that the index variables $k$,$m$ and $n$ are defined to lie in $\mathcal{K}$, $\mathcal{M}$ and $\mathcal{N}$ respectively, unless otherwise stated. The Joint-PPP problem formulation (P1) includes a set of multiplicative fractional expressions involving products of continuous and binary variables (Eqs.\eqref{eq:PForig_MinSINR}, \eqref{eq:BasicSINRexpr}). Hence, (P1) belongs to the class of MINLPs, which are very hard to solve. Their intractability stems from the combination of combinatorial and difficult non-convex non-linear optimization. Usually global optimality guarantees for MINLPs are not provided and in practice even regular-sized problem instances cannot be solved~\cite{LeLe12}. Given that in the UDN framework tens or even hundreds of ANs and UEs may be considered, formulation (P1) cannot be practically further exploited.

\section{Optimal Solution of the PPP Problem through Integer Linear Programming}\label{sec:OPTIMAL_JOINTPPP}

\emph{Motivation:} Due to the intractability of the original MINLP formulation, we seek for methods allowing the acquisition of the exact optimal solution at least for small/medium scale networks. The performance of this solution provides a performance benchmark for any other suboptimal approach. In particular, we will show that the joint PPP problem can be reformulated and exactly solved through a series of Integer Linear Programs (ILPs), without applying any approximation or relaxation. The cost of the linearization reformulation is an increase in the number of variables and constraints, as well as the need to solve multiple optimization problems for locating the common maximum effective SINR. Although ILPs do not admit polynomial-complexity solving algorithms and their worst-case complexity scales exponentially with the problem size, there exists a multitude of powerful algorithms and solvers (such as CPLEX, GUROBI, and MOSEK), which deal with such large-scale problems (involving hundreds or even thousands of variables/constraints), in very fast time-scales, and provide solutions with global optimality guarantees. These features make the problem linearization a well-established method~\cite{ShLi09}.

\emph{Reformulation Procedure:} First, by inspecting the SINR definition in \eqref{eq:BasicSINRexpr}, we observe that for any arbitrary $\left\langle {k,m,n} \right\rangle$ tuple, the allocated power could take a non-zero value if the $k^{\text{th}}$ UE is served by the $m^{\text{th}}$ AP in the $n^{\text{th}}$ partition. This means that we can introduce a new semi-continuous power design variable ${\mathbf{P}} \in {\mathbb{R}^{K \times M \times N}}$, which is a simple extension of the original power variable $\mathbf{p}$ in the AN dimension. In order to guarantee that zero-power is allocated to non-paired AN-UE combinations at any partition, the following additional set of (linear) constraints should be imposed:
\begin{equation}\label{ILP_powsemicontcon}
0 \leqslant {P_{kmn}} \leqslant {p_{\max }} \cdot {\rho _{kmn}},\forall k,\forall m,\forall n.
\end{equation}
Secondly, in order to handle the non-linearity of the effective SINR constraint expression caused by the product of the per-partition SINR contribution, we exploit the fact that each UE is scheduled at exactly one partition, and therefore, on a user basis, only one of the product terms in each $\prod\limits_{n \in \mathcal{N}} {\left( {1 + {\gamma _{kn}}} \right)}$ is non-zero. This behavior can be captured by introducing a new binary variable ${z_{kn}}$ for every possible user/partition combination which indicates if the $k^{\text{th}}$-UE is active at the particular $n^{\text{th}}$-partition. Obviously the new variable is associated with the original pairing and partitioning allocation variable \text{\boldmath{$\rho$}} through the following expression:
\begin{equation}\label{ILP_zvar}
{z_{kn}} = \sum\limits_m {{\rho _{kmn}}} ,\forall k,\forall n,{z_{kn}} \in \left\{ {0,1} \right\},
\end{equation}
since exactly one AN should serve this UE at the particular partition for $z_{kn}$ to be non-zero.
As a third step, we assume a specific (feasible) attainable effective SINR threshold per UE, $\theta_0$. We will later show how to generalize the analysis in order to locate the maximum attainable SINR threshold $\theta^*$. By combining the new power and UE-activity design variables, the fact that the SINR threshold is now a problem parameter and not a variable, and applying a simple terms reordering, the original minimum effective SINR per UE constraints are rewritten as:
\begin{align}\label{ILP_SemiContILPMinSINRcons}
{\theta _0}{z_{kn}} + {\theta _0}\sum\limits_{i \in \mathcal{K}\backslash \left\{ k \right\}} {\sum\limits_{m \in \mathcal{M}} {{g_{km}}{z_{kn}}{P_{imn}}} }  - \nonumber \\ \sum\limits_{m \in \mathcal{M}} {{g_{km}}{P_{kmn}}}  \leqslant 0,\forall k,\forall n.
\end{align}
Under the new formulation, the first and the last group of terms in \eqref{ILP_SemiContILPMinSINRcons} are purely linear, whereas the second group contains non-linear terms, which are formed by products of a binary ($z_{kn}$) and a real bounded variable ($P_{imn} \leq p_{max}$). Such products are amenable to a simple exact linearization procedure, following the approach originally proposed in \cite{Li94} and elaborated in \cite{Wu97}. This comprises the fourth and last step of the linearization reformulation. In particular, we can replace an arbitrary product term ${{z_{kn}}{P_{imn}}}$ with a newly introduced variable ${u_{imnk}}$ (where $i$ represents the potential interfering UE indices with respect to each UE $k$, namely $i \in \mathcal{K}\backslash \left\{ k \right\}$) . We also have to add a set of four (linear) inequality constraints, which guarantee that the original ($\mathbf{z},\mathbf{P}$) and the new ($\mathbf{u}$) variables have identical impact on the optimization model behavior:
\begin{subequations}\label{eq:ILP_AuxConsLinear}
\begin{gather}
  {u_{imnk}} = {z_{kn}}{P_{imn}},\forall k,\forall i \ne k,\forall m,\forall n \hfill \label{eq:ILP_AuxConsLinear_Var}\\
  {P_{imn}} - {u_{imnk}} \leqslant \mathcal{B} \cdot \left( {1 - {z_{kn}}} \right), \hfill \nonumber \\ {u_{imnk}} \leqslant {P_{imn}},{u_{imnk}} \leqslant \mathcal{B} \cdot {z_{kn}},{u_{imnk}} \geqslant 0. \hfill \label{eq:ILP_AuxConsLinear_Ineq}
\end{gather}
\end{subequations}
Note that $\mathcal{B}$ is a big number (readers are referred to the the Big-M method for further details~\cite{ChBa10}) at least equal to the upper bound of the product to be linearized. By simple inspection of constraints in \eqref{eq:ILP_AuxConsLinear} the equivalence of the new modeling variable for the binary-real products is proven. Utilizing the newly introduced variables, constraints \eqref{ILP_SemiContILPMinSINRcons} are now purely linearized:
\begin{align}\label{ILP_LinearMinSINRcons}
{\theta _0}{z_{kn}} + {\theta _0}\sum\limits_{i \in \mathcal{K}\backslash \left\{ k \right\}} {\sum\limits_{m \in \mathcal{M}} {{g_{km}}{u_{imnk}}} }  - \hfill \nonumber \\ \sum\limits_{m \in \mathcal{M}} {{g_{km}}{P_{kmn}}}  \leqslant 0,\forall k,\forall n
\end{align}

\emph{The Joint PPP Solution:} For a target effective SINR threshold per UE $\theta_0$, we have defined a new feasibility optimization problem ${\text{P2}}\left( {{\theta _0}} \right)$, which depends on the adopted threshold and is formulated as an ILP. The complete problem formulation is as follows:
\begin{equation}\label{eq:JointPPP_ILP}
{\text{P2}}\left( {{\theta _0}} \right){\text{: \textbf{find} }} {\mathbf{P}} {\text{ s.t.   \eqref{eq:PForig_Connectivity}, \eqref{ILP_powsemicontcon}, \eqref{ILP_zvar}, \eqref{eq:ILP_AuxConsLinear_Var}, \eqref{eq:ILP_AuxConsLinear_Ineq}, \eqref{ILP_LinearMinSINRcons}}}.
\end{equation}
The formulation given in \eqref{eq:JointPPP_ILP} can be straightforwardly used when known target rate requirements per UE are demanded. In the general case, when target rates are not a-priori known, we search for the maximum common attainable rate and a bisection search \cite{BoVa04} for the optimal threshold is applied. In particular, at each step an effective SINR threshold is considered and then its feasibility is examined by solving problem (P2). The search interval progressively decreases and ultimately converges to the maximum feasible solution. The advantage of the proposed approach is that it converges in a bounded number of iterations given the initial search interval and convergence tolerance~\cite[Sec.4.2]{BoVa04}. Algorithm~\ref{alg:BSJointPPP} (in the following table) summarizes the search for the maximum supported common rate procedure.
\begin{algorithm}
\small
\caption {Optimal Network Coordination by solving the Joint PPP problem}
\begin{algorithmic}[1]
\State Given bounds $\theta_{min} \le \theta_0 \le {\theta_{max}}$, convergence tolerance $\epsilon$
\Repeat
    \State $\theta_0 \leftarrow {{\left( {{\theta_{min}} + {\theta_{max}}} \right)} \mathord{\left/
 {\vphantom {{\left( {{\theta_{min}} + {\theta_{max}}} \right)} 2}} \right.
 \kern-\nulldelimiterspace} 2};$
    \State Solve the ILP feasibility problem ${\text{P2}}\left( {{\theta _0}} \right)$ in \eqref{eq:JointPPP_ILP};
    \State If a feasible solution is found $t_{max} \leftarrow \theta_0$  else $t_{min} \leftarrow \theta_0$;
\Until {$\left| {{\theta_{max}} - {\theta_{min}}} \right| \le \varepsilon$}.
\end{algorithmic}
\label{alg:BSJointPPP}
\end{algorithm}

The linerization reformulation comes with an increase in the problem variables and constraints. The MINLP formulation (P1) in \eqref{eq:PForig} involves $KMN + K$ variables and $3K + KN + MN$ constraints. Assuming that there is an almost equal number of ANs and UEs ($M \approx K$), which is also significantly larger than the number of partitions ($M\gg N$), after some mathematical manipulations, we can prove that the number of variables scales quadratically while the number of constraints scales linearly with respect to the number of ANs $M$ and UEs $K$. As far as the ILP problem formulation (P2) in \eqref{eq:JointPPP_ILP} is concerned, the number of variables equals $2KMN + KN + \left( {K - 1} \right)KMN$ and the number of constraints equals $3KN + MN + K + KMN + 5K\left( {K - 1} \right)MN$. After simple mathematical manipulations, a third order scaling can be shown for the number of variables and constraints with respect to $M$ or $K$.

\emph{The Fixed Pairing Case:} We also briefly examine the special case of fixed pairing, which draws our attention due to two reasons. First, by constraining UEs to be associated with certain ANs, we significantly reduce solution complexity, since the original joint PPP problem resorts to the problem of partitioning and determining the power of the various AN-UE pairs. Secondly, there exist various fixed pairing strategies, which are widely applicable to practical systems and may not be desirable to change. Typical pairing policies correspond to the ``best AN per UE selection", based on criteria such as distance, received signal strength, received SINR, biased SINR, etc.~\cite{An13,GoAl13}. By fixing pairing, optimal coordination is greatly simplified: the 3D pairing/partitioning and power design variable matrices in \eqref{ILP_powsemicontcon} reduce to 2D matrices dimensioned as $K \times N$, whereas the minimum SINR guarantees become ${\theta _0}{\rho _{kn}} - {{\tilde g}_{kk}}{P_{kn}} + {\theta _0}\sum\limits_{i \in \mathcal{K}\backslash \left\{ k \right\}} {{{\tilde g}_{ki}}\left( {{\rho _{kn}}{P_{in}}} \right)}  \leqslant 0,\forall k,\forall n, $ and can be linearized straightforwardly as before. Note that for an arbitrary UEs pair $\left\langle {i,j} \right\rangle$, ${{{\tilde g}_{ij}}}$ denotes the channel gain between the $i^{th}$ UE and either its serving AN (for $i=j$) or the serving AN of the $j^{th}$ UE (for $i \ne j$). Therefore, the reduced complexity partitioning and power coordination problem given fixed pairing is also solved through the ILP framework. It is also easy to show that both the number of variables and the number of constraints scale quadratically with the network size (number of ANs and UEs), as opposed to the third-order scaling of the joint PPP problem.

\section{Suboptimal Solutions for the PPP problem}\label{sec:SuboptimalPPP}

\subsection{Motivation and Design}\label{sec:SuboptimalPPP_Intro}
The major difficulties when coping with the problem in \eqref{eq:JointPPP_ILP} stem from the joint handling of all design variables (partitioning, pairing, and power allocation), as well as the discrete/combinatorial nature of pairing and partitioning resulting in exponential complexity scaling. Applying the optimal ILP model is therefore realistic for limited network sizes (up to 10-20 ANs and UEs as it has been observed through typical computational experiments). Instead, for larger UDNs comprising tens or even hundreds of serving and served nodes, lower-complexity yet close-to-optimal solutions are required; this is the objective of the study performed in this Section.

In order to relax joint variables consideration, we examine the components of the joint PPP problem, namely pairing, partitioning and power coordination. Regarding \emph{pairing} the strategy of assigning the closest available AN to each UE is reasonable; this is justified by the fact that direct link path gain (between a UE and its serving AN) is maximized, while at the same time power coordination allows for limiting the interference caused to the rest of UEs (this is also validated through simulations later). As far as the \emph{partitioning} problem component is concerned, its combinatorial nature is tackled by use of a greedy UE-to-partition assignment approach, which optimizes an adopted objective function (such as the common rate or the network interference) at each local stage, leading to ``good" (but not necessarily globally optimal) final assignments, followed by an optional refinement procedure, which exchanges AN-UE pairs between already assigned partitions. Finally, given that a set of mutually-exclusive AN-UE pairs have been already partitioned, optimal \emph{power coordination} is readily available~\cite{TaCh13}.
\begin{figure}[]
\centering
\includegraphics[scale=0.50]{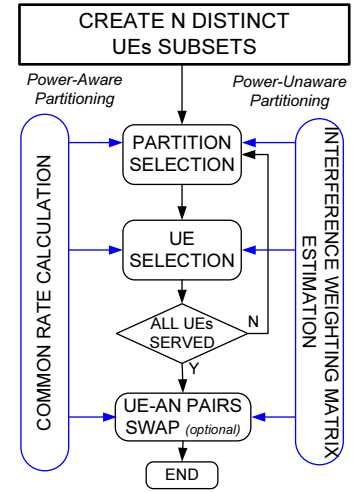}
\caption{The algorithmic framework for obtaining suboptimal PPP solutions}
\label{fig:SuboptimalPPPframework}
\end{figure}
Fig.\ref{fig:SuboptimalPPPframework} provides an overview of the proposed framework, along with the two algorithmic variations. Key to this framework is how to carefully select the coexisting sets of AN-UE pairs (namely the AN-UE pairs served at each partition) since spatial reuse of resources and network interference levels are directly impacted~\cite{ToHa12}. For each algorithmic variation we provide a high-level description of the partitioning strategy, that is how to select the best UE/partition pair at each iteration, and also provide implementation details regarding the selection criteria (rate maximization, interference minimization) and the associated metrics estimation (exact or approximated). Note that partitioning decisions are strongly coupled with the applied pairing and power coordination strategy.

\subsection{Power-Aware Partitioning}\label{SuboptimalPPP_PowAwarePartitioning}
 We assume $K$ UEs and $N$ partitions, where $K>N$, and we provide the necessary additional notation. Let an arbitrary network partition $n$, and $\mathcal{L}_n$ the assigned UEs subset (comprising $L_n$ elements), for which ${\mathcal{L}_n} \subseteq K,{\mathcal{L}_n} \cap {\mathcal{L}_{n'}} = \emptyset ,n \ne n'$. For a given UEs pair $\left\langle {i,j} \right\rangle$ belonging to the partition $\mathcal{L}_n$ we remind that ${{{\tilde g}_{ij}}}$ denotes (for the $i^{th}$ UE) either the direct link channel gain (that is towards the serving AN if $i=j$) or the interfering channel gain (towards the serving AN of the $j^{th}$ UE, if $i \ne j$). Clearly, this information could be stored in a square matrix $\mathbf{\tilde G}$ with dimension $L_n \times L_n$, or could be split into an $L_n$-size vector $\mathbf{v}$ containing the inverse of the main diagonal elements and a square $L_n \times L_n$ matrix $\mathbf{F}$ including the remaining elements and zeros in the main diagonal~\cite{TaCh13}. Finally let ${{r^*}\left( {{\mathcal{L}_N}} \right)}$ an estimate of the achieved common rate (details on the calculation of $r^*$ will be given right after the partitioning approach is described).
\subsubsection{Partitioning}
 We now describe the greedy allocation procedure. Since partitioning aims at maximizing the achieved rate, we initially select the $N$ best --in terms of experienced channel gain-- UEs maximizing the received signal power, and assign each of them to one of the $N$ orthogonal partitions. Clearly, such a strategy corresponds to a performance upper bound since intra-partition interference is eliminated but not all UEs are served. If $i$ the allocation cycle index, $\left\langle {{n^*}\left( i \right),{k^*}\left( i \right)} \right\rangle$ the selected partition and UE indices, $\mathcal{S}$ the pool of yet unassigned UEs ($\mathcal{S}$ initialized to $\mathcal{K}$), and $\mathcal{M}_n$ the set of available ANs in partition $n$ (all $\mathcal{M}_n$ sets initialized to $\mathcal{M}$), then the first $N$ cycles are performed as follows:
\[\begin{gathered}
\small
  {\text{for }}i{\text{  =  }}1{\text{:}}N \hfill \\
  {\text{     }}\left| \begin{gathered}
  {\text{Partition Selection:  }} {n^*}\left( i \right) = i; \hfill \\
  {\text{Pair Selection: }} \left\langle {{k^*}\left( i \right),{m^*}\left( i \right)} \right\rangle  = \operatorname{argmax} \left\{ {{g_{km}}} \right\}_{k \in \mathcal{S}}^{m \in {\mathcal{M}_n}} \hfill \\
\end{gathered}  \right. \hfill \\
\end{gathered} \]

We continue by gradually allocating the remaining $K-N$ UEs to the orthogonal partitions so as the reduction in the achieved rate due to the intra-partition interference is kept low. To this end, a UE-partition combination is selected at each algorithmic iteration as follows. First among the $N$ partitions, the one which experiences the maximum rate so-far is chosen. This is justified by the fact that this partition experiences the lowest interference, and since fair rates should be assigned among all UEs, the total traffic load should be balanced among the available orthogonal resources. Secondly, for every element of the unserved UEs pool, the updated common rate of the selected partition is calculated. The UE that adds the minimum possible extra interference and thus results to the maximum updated partition rate is selected and assigned to this partition. Then the supported rates per partition are updated and, if the pool is not empty, the search for new UE-partition combination starts again. The procedure is summarized below:
\[\begin{gathered}
\small
  {\text{for }}i{\text{  =  }}N + 1{\text{:}}K \hfill \\
  {\text{     }}\left| \begin{gathered}
  {\text{Partition Selection: }}{n^*}\left( i \right) = \arg \mathop {\max }\limits_n r\left( {{\mathcal{L}_n}} \right) \hfill \\
  {\text{UE Selection:  }}\forall k \in \mathcal{S}:{{r'}_k} = {r^*}\left( {{\mathcal{L}_{{n^*}\left( i \right)}} \cup k} \right), \\{k^*}\left( i \right) = \arg \mathop {\max }\limits_{k \in S} \left\{ {{{r'}_k}} \right\} \hfill \\
  {\text{Update UEs pool:  }}\mathcal{S} = \mathcal{S}\backslash \left\{ {{k^*}\left( i \right)} \right\} \hfill \\
\end{gathered}  \right. \hfill \\
\end{gathered} \]
An (optional) partitioning refinement follows. The fact that the UE-to-partition assignment is performed in multiple independent steps and not at a single-shot, provides probably good local solutions but no guarantee for global optimality (as any greedy approach). To improve the final solution we propose an additional reallocation phase, which searches for further performance improvement at the expense of increased complexity. The refinement procedure consists of a set of iterations. At each iteration we search for AN-UE pair exchanges among different partitions, which if applied, result to a rate increase. For two arbitrary partitions $n,{n'}$ where $n \ne n'$ and a pair of UEs $\left\langle {k,k'} \right\rangle $ where $k \in {\mathcal{L}_n},k' \in {\mathcal{L}_{n'}}$, their contribution to the system rate is $r_{before}^* = {r^*}\left( {{\mathcal{L}_n}} \right) + {r^*}\left( {{\mathcal{L}_{n'}}} \right)$, whereas after swap becomes: $r_{swap}^* = {r^*}\left( {\left( {{\mathcal{L}_n}\backslash \left\{ k \right\}} \right) \cup k'} \right) + {r^*}\left( {\left( {{\mathcal{L}_{n'}}\backslash \left\{ {k'} \right\}} \right) \cup k} \right)$. If a positive rate difference, $\Delta r = r_{swap}^* - r_{before}^* > 0$, is estimated, the particular UEs are re-assigned to the examined partitions. One may strike a balance between performance improvement and additional complexity, by limiting the number of explored pair exchanges. In our implementation we perform a single sweep of the $N$ partitions searching for potential swaps until no further rate improvement is possible. Partition formation is followed by a per-partition power coordination. Algorithm~\ref{alg:ApproxSol1} given in Appendix~\ref{sec:Appendix1} provides the complete pseudocode description.

\emph{To sum-up, we have proposed a low-complexity UEs partitioning scheme given an estimate of the achieved common rate for each UEs-subset to partition assignment.}

\subsubsection{(Prerequisites) Common rate estimation given a UEs-subset partitioning}\label{SuboptimalPPP_PairingAndPowerCoord} Common rate estimation calls for solving the joint pairing and power coordination problem, which is again a difficult MINLP problem. Since the focus of the presented solution is low-complexity we follow a 2-step disjoint approach, where we first perform pairing and then apply power coordination considering the formulated AN-UE pairs. As far as \emph{pairing} is concerned, the following Rule is applied:
\begin{lem}\label{lem:Pairing}
Each UE (indexed by $k$) is associated with the best available AN (indexed by ${m^*}\left( k \right)$) in terms of experienced channel gain, that is, $\forall k \in \mathcal{L}_n:{m^*}\left( k \right) = \arg \max {\left\{ {{g_{km}}} \right\}_{m \in \mathcal{M}}}$, where ties are resolved by prioritizing UEs with highest channel-gains (hence a UE may not be associated with its closest AN).
\end{lem}
After the pairs have been formed, per-pair \emph{power} allocation can be found optimally according to \cite{TaCh13}. In particular, the authors in \cite{TaCh13} used the Perron-Frobenius framework \cite{BePl94} for acquiring the optimal common SINR for an $\mathcal{L}_n$-pairs network and the resulting power solution vector. Utilizing the alternative channel representation given by matrix $\mathbf{F}$ and vector $\mathbf{v}$  (as described above), they provided the following expression for obtaining the exact common SINR ${\gamma^*\left( {{\mathcal{L}_n}} \right)}$:
\begin{equation}\label{eq:CommonSINROptExact}
{\left( {\gamma^*\left( {{\mathcal{L}_n}} \right)} \right)^{ - 1}} = \mathop {\max }\limits_{1 \leqslant i \leqslant {L_n}} \rho \left( {{\mathbf{F}}\left( {{\mathcal{L}_n}} \right) + \left( {{1 \mathord{\left/
 {\vphantom {1 {{p_{\max }}}}} \right.
 \kern-\nulldelimiterspace} {{p_{\max }}}}} \right) \cdot {\mathbf{v}}\left( {{\mathcal{L}_n}} \right){\mathbf{e}}_i^T} \right),
 \end{equation}
where $\rho \left(  \cdot  \right)$ is the largest eigenvalue of a non-negative matrix or simply the Perron root of it~\cite{BePl94} and $\mathbf{e}_i$ the $i^{\text{th}}$ unit coordinate vector. Then, the achieved common rate is acquired by applying the Shannon-rate formula, as summarized in the following Rule:
\begin{lem}\label{lem:CommonRateExact}
For a set of AN-UE pairs ${\mathcal{L}_n}$ assigned to a partition $n$ the achieved maximum common rate is exactly calculated as ${r^*}\left( {{\mathcal{L}_n}} \right) = \left( {{1 \mathord{\left/
 {\vphantom {1 N}} \right.
 \kern-\nulldelimiterspace} N}} \right) \cdot {\log _2}\left( {1 + {\gamma ^*}\left( {{\mathcal{L}_n}} \right)} \right)$, where ${{\gamma ^*}\left( {{\mathcal{L}_n}} \right)}$ is given from \eqref{eq:CommonSINROptExact}.
\end{lem}
Although \eqref{eq:CommonSINROptExact} is straightforward to use, it involves the calculation of the maximum largest eigenvalue of $L_n$ matrices of size $L_n \times L_n$ each. Since we focus on dense network deployments, tens or even hundreds of paired AN-UE connections would coexist, leading to a non-negligible complexity for calculating the required eigenvalues. To reduce the aforementioned complexity we propose a series of approximations. We observe that in \eqref{eq:CommonSINROptExact} we compare the spectral radius of $L_n$ matrices in order to identify the maximum one. These $L_n$ matrices differ only in one column. A first approximation corresponds to the consideration of a single matix, that is the one formed by the maximum $v_i$ element (equivalently the worst direct link channel gain over all pairs, since $\mathbf{v}$ contains the inverse direct channel gains). If $l$ is this pair index, then the only matrix to be examined from the complete $L_n$ matrices-set is given in \eqref{eq:SINRboundStep1}, where we have dropped the $\mathcal{L}_n$ identifier, since we refer to this arbitrarily formed partition:
\begin{equation}\label{eq:SINRboundStep1}
{\mathbf{A}} = {\mathbf{F}} + \left( {{1 \mathord{\left/
 {\vphantom {1 {{p_{\max }}}}} \right.
 \kern-\nulldelimiterspace} {{p_{\max }}}}} \right) \cdot {\mathbf{ve}}_l^T{\text{, for }}l = \arg \mathop {\max }\limits_{1 \leqslant i \leqslant {L_n}} {\left[ {\mathbf{v}} \right]_i}.
\end{equation}
A second approximation step comprises the replacement of the exact largest eigenvalue calculation (for matrix ${\mathbf{A}}$) with simpler approximation expressions based on developed Perron-root bounds found in the literature (refer to~\cite{Liu96} for a review of Perron-root bounds). The bounds calculation is based on expressions involving simple row-sums and column-sums of the matrix and its powers. In our paper we utilize the bounds proposed in \cite{ChLi07} which strike a balance among accuracy and complexity. We now summarize the method whereas complete details regarding the extraction of these bounds can be found in \cite{ChLi07}.

For a square matrix $\mathbf{Z}$ with $L$ rows/columns, the row-sums and column-sums of the $l^{th}$ row and column correspondingly are defined as $r{s_l}\left( {\mathbf{Z}} \right) = \sum\limits_{1 \leqslant j \leqslant L_n} {{z_{lj}}} ,c{s_l}\left( {\mathbf{Z}} \right) = \sum\limits_{1 \leqslant j \leqslant L_n} {{z_{jl}}}.$
Then, for the matrix of interest $\mathbf{A}$ a new matrix (let $\mathbf{B}$) polynomially related to $\mathbf{A}$ is introduced. There is no single selection option for $\mathbf{B}$; in \cite{ChLi07} ${\mathbf{B}} = {\left( {{\mathbf{A}} + {\mathbf{I}}} \right)^{L_n - 1}}$ is proposed. Given ${\mathbf{A}}$ and ${\mathbf{B}}$ and $a,b$ integer parameters (typically equal to 1 or 2) the following quantities are defined for each row/column $l$:
\begin{equation}\label{eq:PerronBounds}
{\delta _l} = {\left( {\frac{{r{s_l}\left( {{A^a}{B^b}} \right)}}{{r{s_l}\left( {{B^b}} \right)}}} \right)^{1/a}},{{\delta '}_l} = {\left( {\frac{{c{s_l}\left( {{A^a}{B^b}} \right)}}{{c{s_l}\left( {{B^b}} \right)}}} \right)^{1/a}}.
\end{equation}
Based on those expressions we get the following bounds for the inverse common SINR ${\left( {{\gamma ^*}} \right)^{ - 1}}$:
\begin{equation}\label{eq:SINRbounds_rowcol}
\begin{gathered}
  \left( {{\gamma ^*}} \right)_{lbrs}^{ - 1} = \mathop {\min }\limits_l {\delta _l} \leqslant {\left( {{\gamma ^*}} \right)^{ - 1}} \leqslant \mathop {\max }\limits_l {\delta _l} = \left( {{\gamma ^*}} \right)_{ubrs}^{ - 1}, \hfill \\
  \left( {{\gamma ^*}} \right)_{lbcs}^{ - 1} = \mathop {\min }\limits_l {{\delta '}_l} \leqslant {\left( {{\gamma ^*}} \right)^{ - 1}} \leqslant \mathop {\max }\limits_l {{\delta '}_l} = \left( {{\gamma ^*}} \right)_{ubcs}^{ - 1} \hfill. \\
\end{gathered}
\end{equation}
Exploiting both row-wise and column-wise expressions, we bound the SINR by:
\begin{align}
\gamma _{LB}^* = \max \left\{ {\left( {{\gamma ^*}} \right)_{ubrs}^{ - 1},\left( {{\gamma ^*}} \right)_{ubcs}^{ - 1}} \right\}, \nonumber \\
\gamma _{UB}^* = \min \left\{ {\left( {{\gamma ^*}} \right)_{lbrs}^{ - 1},\left( {{\gamma ^*}} \right)_{lbcs}^{ - 1}} \right\}.
\label{eq:SINRbounds_overall}
\end{align}
Finally by assuming that the approximated SINR is in the middle of the interval defined by $\gamma _{LB}^*$ and $\gamma _{UB}^*$, we resort to the following Rule:
\begin{lem}\label{lem:PerronBounds}
For a set of AN-UE pairs ${\mathcal{L}_n}$ assigned to a partition $n$, the achieved maximum common rate is approximately calculated as $r_{{\text{approx}}}^*\left( {{\mathcal{L}_n}} \right) = \left( {{1 \mathord{\left/
 {\vphantom {1 N}} \right.
 \kern-\nulldelimiterspace} N}} \right) \cdot {\log _2}\left( {1 + \gamma _{{\text{approx}}}^*\left( {{\mathcal{L}_n}} \right)} \right)$, where $\gamma _{{\text{approx}}}^*\left( {{\mathcal{L}_n}} \right) = \left( {{1 \mathord{\left/
 {\vphantom {1 2}} \right.
 \kern-\nulldelimiterspace} 2}} \right) \cdot \left( {\gamma _{LB}^* + \gamma _{UB}^*} \right)$ is the approximated common SINR, obtained in turns by applying \eqref{eq:PerronBounds},\eqref{eq:SINRbounds_rowcol},\eqref{eq:SINRbounds_overall} to the matrix defined in \eqref{eq:SINRboundStep1}.
\end{lem}
\emph{To sum up, we have provided simple rules and expressions for calculating (either exactly or approximately) the achieved common rate for any UE subset allocated to an arbitrary partition. These results can be leveraged in any \emph{partitioning} procedure.}

\subsection{Power-Unaware Partitioning}\label{SuboptimalPPP_PowUnawarePartitioning}
An alternative lower-complexity partitioning scheme is also provided. The complexity of the previous approach is dominated by the achieved rate calculations. These calculations need to take into account both the channel gain matrix and the dynamically tuned power per pair in order to provide reliable SINR estimates. Accounting for power dynamics is highly important, since a UE residing close to its serving AN contributes less to the overall network interference than a UE residing farther. Here, we ignore the power dynamics and work towards isolating as much as possible strongly interfering UEs, given a limited number of orthogonal partitions, leveraging only the direct- and cross-links channel gain information. This comes with a decrease in the partitioning formation complexity but also with a suboptimality cost. The proposed solution consists of two steps, first the creation of a UE-to-UE Interference Weighting Matrix, let $\mathbf{E}$, and then a greedy allocation algorithm which aims at minimizing the overall network interference based on the estimated interference levels\footnote{Note the resemblance of the particular modeling and solution with the graph-based framework and the max-k-cut partitioning problem, refer for example to \cite{ChRa93} and \cite{ChTa09}.}.

\subsubsection{Interference Weighting Matrix Estimation} We form a symmetric $K \times K$ matrix $\mathbf{E}$ for which its arbitrary element $e_{ij}$ expresses an estimation of the interference caused to the $j^{\text{th}}$-UE by serving the $i^{\text{th}}$-UE in the same partition. Therefore, a large $e_{ij}$ entry calls for assigning UEs $i$ and $j$ in different partitions, whereas a zero or small entry implicates reuse of the same resources by the UEs without severely hurting system performance. For each UE a set of dominant interferers (without considering UEs served by the same AN) is assumed and typically accounting for 2-3 dominant interferers is enough. Then the interference weight $e_{ij}$ for an arbitrary UEs pair $\left\langle {i,j} \right\rangle$ where $m$ is the serving AN of UE $i$ is calculated as follows:
\begin{itemize}
  \item If $i$ and $j$ are served by the same AN, set $e_{ij}$ to infinity (practically to a very large number, e.g. $10^5$) since intra-AN UEs should not share the same partition;
  \item If $i$ does not belong to the dominant interfering set of $j$, set $e_{ij}$ to zero, since they can share the same partition and reuse the same resources causing limited increase to the network interference levels;
  \item If $i$ belongs to the dominant interfering set of $j$, set ${e_{ij}} = {{\tilde g}_{jm}}$, considering that interference intensity is proportional to the signal path-gain towards the victim UE.
\end{itemize}

\subsubsection{Partition and UE Pairs Selection for Minimizing Network Interference}
The objective of partitioning is to minimize the overall network interference using the information stored in the weighting matrix $\mathbf{E}$. This can be viewed as a combinatorial assignment problem, hence here we propose a greedy allocation approach as well. The approach involves $K$ allocation cycles, where at each cycle an unassigned UE is first selected and then the partition to allocate the particular UE is chosen. This is done as follows: i) UEs experiencing the worst estimated interference conditions (by considering the dominant interfering signals) from the unassigned UEs pool are promoted. By selecting first the UEs with the worse estimated SINR conditions, we create assignments protecting them as much as possible, since at each subsequent assignment step the interference caused to the most impaired UEs is taken into account; ii) Then the algorithm selects the partition to which the smallest increase of the sum of weights is caused if the selected UE is assigned to it. Following this strategy we expect to cause the smallest interference increase and accordingly rate decrease to network partitions. \emph{To sum up, we proposed a greedy partitioning approach applying common principles with the power-aware partitioning algorithm of Section~\ref{SuboptimalPPP_PowAwarePartitioning}, but with significantly reduced complexity since it does not take into account the power dynamics of different AN-UE pairs.}

\section{Results \& Discussion}\label{sec:Results}

\subsection{Simulation Setup and Algorithms Overview}\label{sec:SimulationSetup}
We now discuss performance evaluation and parametrization aspects of UDNs as extracted from computer simulation experiments, focusing on: i) justifying the need for across the network coordination and evaluating the achieved network rates by applying the proposed sophisticated pairing, partitioning and power control mechanisms, ii) estimating the quality and suboptimality penalties of the proposed lower-complexity solutions, and iii) exploring the impact of different system parameters (network size, partitioning size, relative density of AN to UE population) on network performance and best coordination policy.

%

For each scenario 1000 independent network setups are simulated, where ANs and UEs are randomly and uniformly dropped over a square area covering $1 \text{ km}^2$ similar to a typical macro-cell. We examine various AN and UE density scenarios (denoted by $\lambda_{AN}$ and $\lambda_{UE}$ respectively) by varying the number of ANs and UEs in the same geographic area. A log-distance large-scale propagation model with path-loss exponent equal to $4$ is applied for both useful and interfering signal transmissions. Each paired AN-UE transmission is limited by a $30 \text{ dBm}$ maximum power budget ($p_{max}$). At each receiver, thermal noise of density $-174 \text { dBm/Hz}$ is considered. As far as the Integer Linear Programming problems, we model them using the {CVX} framework~\cite{CVX} and solve them through the {GUROBI} optimizer~\cite{GUROBI}. As far as the partitioning size $N$, we distinguish between 2 approaches: i) ``fixed-$N$, where a particular constant partitioning size is considered for each network realization, and ii) ``dynamic-$N$", where the size is dynamically tuned at each realization according to a selected policy. For example, $N$ could be selected such that all UEs selected to be served by an AN are orthogonalized. This is accomplished if at each realization $N$ is set equal to the maximum number of UEs paired with a single AN. Before presenting the results, an overview of the proposed coordination algorithmic solutions and the two adopted baseline approaches, is presented in Table I.

\begin{table*}[]
\renewcommand{\arraystretch}{1.3}
\centering
\begin{threeparttable}[]
\caption{UDN Network Coordination Approaches}
\begin{tabular}{| L{3.2cm} | L{10cm} | L{2.0cm} |}
\hline
\bfseries Algorithm & \bfseries Short Description & \bfseries Tools/Approach \\
\hline\hline
\textbf{JOINT-PPP} & (Sec.\ref{sec:OPTIMAL_JOINTPPP}) Provides the optimal network coordination by jointly considering pairing, partitioning and power coordination.  & Integer Linear Programming\\
\hline
\textbf{JOINT-PP (Fixed Pairing)} & (Sec.\ref{sec:OPTIMAL_JOINTPPP}) Provides the optimal joint partitioning and power coordination for known pairing (e.g. serving each UE by its closest AN) & Integer Linear Programming\\
\hline
\textbf{SUBOPT-PPP-PowAware / INTELLIGENT-PPP} & (Sec.\ref{SuboptimalPPP_PowAwarePartitioning}) Provides a suboptimal network performance by decoupling pairing/power coordination and partitioning. Power-aware partitioning is applied; two versions are supported depending on the rate calculations method, ``EXACT" for Rule~\ref{lem:CommonRateExact} or ``APPROX" for Rule~\ref{lem:PerronBounds}. The latter is significantly less complex than the former since it avoids multiple eigenvalue calculations of large dense matrices and it is also referred as ``INTELLIGENT-PPP". & Greedy Searches, Eigenvalue Calculations\\
\hline
\textbf{SUBOPT-PPP-PowerUnaware} & (Sec.\ref{SuboptimalPPP_PowUnawarePartitioning}) Provides a suboptimal network performance by fully decoupling pairing, partitioning, and power coordination. A greedy-search partitioning aiming at minimizing the overall estimated network interference without considering power dynamics is applied. After partitioning is completed dynamic power coordination of each partition is employed. & Greedy Searches\\
\hline
\textbf{FULL SPATIAL REUSE} & Corresponds to the simultaneous serving of all UEs over the whole bandwidth. It can be considered as a special case of the PPP problem, with $N=1$. Each UE is associated with its closest AN, followed by optimal power coordination. & Greedy Search, Linear Programming \\
\hline
\textbf{FULL ORTHOGONALIZATION} & The other extreme of the Full Spatial Reuse scheme, providing interference-free transmissions, since each UE is assigned to a different partition (in an FDMA or TDMA fashion) at the expense of bandwidth reduction (FDMA) or medium-access rate (TDMA). & -- \\
\hline
\end{tabular}
\end{threeparttable}
\label{ta:Algorithms}
\end{table*}

\subsection{Simulation Results and Discussion}\label{sec:SimulationResults}
\begin{figure}[]
\centering
\includegraphics[width=0.5\textwidth]{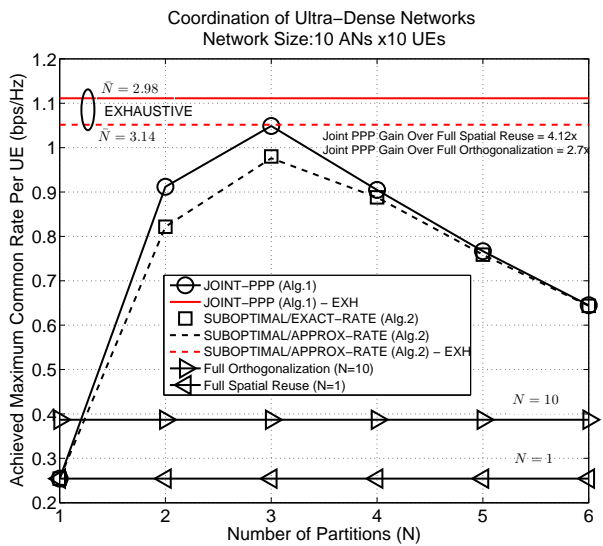}
\caption{Joint and Suboptimal PPP performance for different partitioning in a $10\times10$ network deployment. $\bar N$ stands for the averaged optimal partitioning size per realization acquired through exhaustive search.}
\label{fig:PPP_10x10}
\end{figure}

First, a $10\times10$ network is considered and the maximum achieved common rate for various partitioning scenarios is depicted in Fig.\ref{fig:PPP_10x10}. We observe that the sophisticated network coordination provides significantly enhanced user rates compared to the baseline approaches, thus justifying its additional complexity. In particular, when applying \emph{joint optimal coordination} an almost \emph{4x increase} is observed \emph{compared to full-orthogonalization}, whereas a \emph{2.7x improvement} is achieved \emph{compared to full-spatial-reuse}. Also, \emph{The selection of the optimal partitioning size ($N$) is shown to be highly critical}, since the optimal trade-off point between SINR increase and resources reuse factor decrease should be identified~\cite{JiAn08} (for this setup optimal $N$ is 3). Moreover, both variations of the Power-Aware Suboptimal-PPP algorithm perform very close to the exact optimal; at the optimal partitioning scenario the suboptimal algorithm achieves almost 93\% of the optimal rate. We will consider only the approximated version of the Power-Aware Suboptimal-PPP approach (termed ``Intelligent-PPP") hereafter. In addition, both for the Joint-PPP and the Suboptimal-PPP approaches we present the achieved rates when for every network realization the partition size can be dynamically selected (``optimal $N$ per snapshot) through exhaustive search. These figures provide a performance upper bound, when \emph{dynamic orthogonalization-degree tuning optimization depending on the distribution of UEs and ANs} is allowed, and are shown to agree on average with the $N=3$ case.

\begin{figure*}
\centering
\subfigure[Network Sum-Rate]{%
\includegraphics[scale=.35]{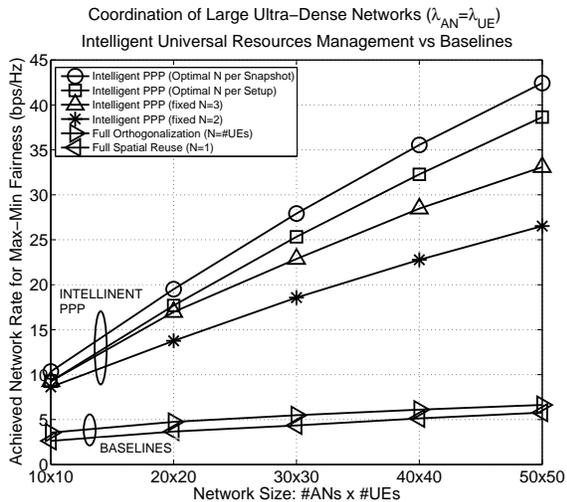}}
\label{fig:RateScaling_SquareNetworks_SumRate}
\subfigure[Maximum Common Rate Per-UE]{%
\includegraphics[scale=.35]{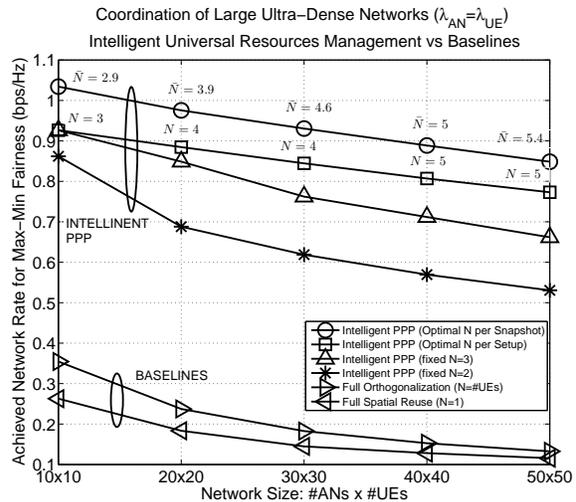}}
\label{fig:RateScaling_SquareNetworks_PerUE}
\caption{Coordinating Ever-Densified Networks: Per-UE and Network Sum-Rate Scaling for $\lambda_{AN}/\lambda_{UE} = 1$.}
\label{fig:RateScaling_SquareNetworks}
\end{figure*}

In Fig.\ref{fig:RateScaling_SquareNetworks}, we illustrate how network performance scales with increasing UE density, while we increase accordingly the infrastructure density, such that $\lambda_{AN}=\lambda_{UE}$. We illustrate the performance and the optimal setting of $N$ (Fig.\ref{fig:RateScaling_SquareNetworks_PerUE}) for the Intelligent-PPP solution regarding both fixed and dynamic partitioning tuning. We observe that, \emph{with an increasing UE density, optimal partitioning size increases }(from $N\approx3$ in the $10\times10$ setup goes up to $N\approx5$ in the $50\times50$ case) since interference conditions worsen.  Thus, the importance of increasing the number of partitions as AN-UE density increases is highlighted. A linear scaling of the network sum-rate is observed for all the intelligent-PPP solutions (Fig.\ref{fig:RateScaling_SquareNetworks_SumRate}), however, the per-UE rate decreases (Fig.\ref{fig:RateScaling_SquareNetworks_PerUE}). Hence, \emph{when strict per-UE rate requirements are imposed, the offered rates cannot be sustained by a $1:1$ infrastructure/UE network densification}.

\begin{figure}
\centering
\includegraphics[width=0.5\textwidth]{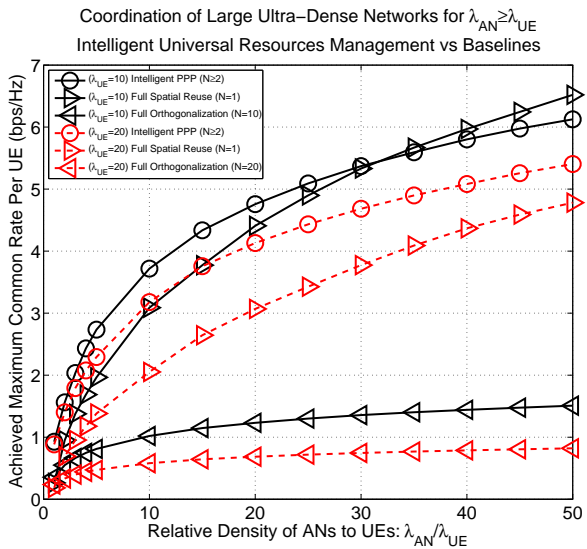}
\caption{Coordinating Ever-Densified Networks: Per-UE and Network Sum-Rate Scaling for $\lambda_{AN}/\lambda_{UE} \geq 1$ and Dynamic Partitioning Size Selection.}
\label{fig:RateScaling_NonSquareNetworks}
\end{figure}

Furthermore in Fig.\ref{fig:RateScaling_NonSquareNetworks} we explore the per-UE rate scaling while increasing the ratio of infrastructure to users density. To this end we keep the number of UEs fixed ($10$,$20$ UEs/$1\text{ km}^2$) and increase the number of access nodes so as to achieve $\lambda_{AN}/\lambda_{UE}$ ratios spanning from $1$ up to $50$. By increasing the relative density of ANs to UEs, the probability for a UE to be served by a closer AN increases. This allows the per-AN power to be decreased, and thus the overall network power to be reduced, leading to increased achieved rates across the network. The \emph{``full-orthogonalization" approach seems to be little affected by the plethora of serving AN options, since bandwidth limitation dominates} (system is in the bandwidth-limited region, where further increase in received power cannot lead to significant increase in rate). This is not the case for the dynamic intelligent-PPP solution, which tunes the partition size so as to achieve the optimum SINR/banwidth trade-off point. The ``full-spatial-reuse" scheme performance is also positively affected by the increased infrastructure density. It is observed that the minimum $\lambda_{AN}/\lambda_{UE}$ density ratio to support a required common per-UE rate is significantly lower when partitioning is applied, compared to full spatial reuse. Interestingly, \emph{for ever increasing infrastructure density } ($\lambda_{AN}/\lambda_{UE}\geq30$ for the $10$-UEs scenario), this is not the case, since full \emph{spatial reuse  outperforms any partitioning scheme }(assuming that ANs not serving any UE are turned-off). At such extreme AN densification scenarios close to interference-free regions are created around each served UE.

\begin{figure}[]
\centering
\includegraphics[width=0.5\textwidth]{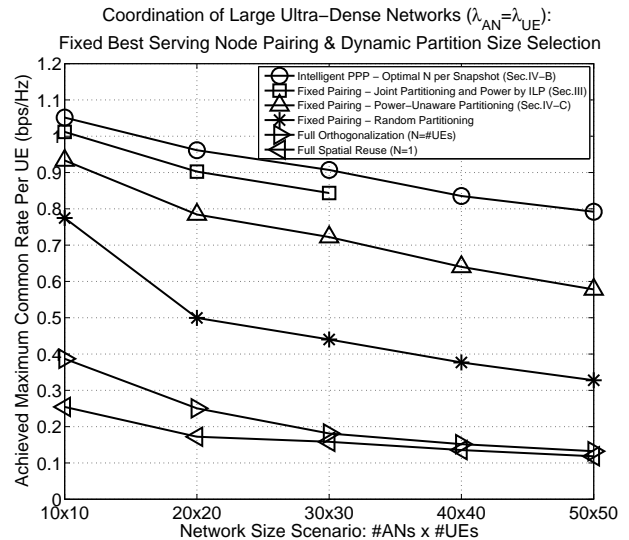}
\caption{Lower Complexity Coordination: Fixed Cellular-paradigm Pairing and Dynamic Partitioning Size Selection for Intra-AN orthogonalization (Joint Partitioning and Pairing solution results are presented up to 30x30 network sizes due to high ILP complexity).}
\label{fig:CellParadigmDynamicN}
\end{figure}

In the final set of experiments we explore a lower complexity coordination strategy which i) fixes pairing according to the ``cellular paradigm", that is, all UEs are served by their closest AN, and ii) dynamically selects the partition size $N$ in order to orthogonalize all AN-UE pairs with the same AN-end. As a reference scheme we utilize the Intelligent-PPP scheme previously shown to perform close-to-optimal, yet admitting relatively low complexity. Indicative performance results are depicted in Fig.\ref{fig:CellParadigmDynamicN} for five network sizes. The joint partitioning and power coordination benchmark for the particular fixed pairing strategy (acquired through ILP) provides the maximum achievable performance, thus allowing us to estimate the penalty of this suboptimal coordination approach. Based on Fig.\ref{fig:CellParadigmDynamicN}, \emph{the performance penalty due to fixed pairing and non-exhaustive search for optimal partitioning size seems to be quite low} ($3.7$--$7$\%). We also examine\emph{ the performance of the 3-step fully decoupled power-unaware partitioning solution} for the current strategy. This solution a\emph{chieves high performance} ($86$--$92$\% of the ILP) with significantly reduced complexity \emph{and also outperforms a baseline random partitioning} by a factor of $1.2$--$1.6$. This is justified by the intelligence of the greedy partitioning procedure, which tends to orthogonalize UEs served by neighboring ANs. Conclusively, it is a good candidate for implementation in large UDNs due to its simplicity and partially distributed nature. Pairing and power coordination can be applied in a fully distributed way, whereas only partitioning requires centralization (due to the interference weighting matrix handling).

\section{Conclusion - Future Research Directions}\label{sec:Conclusion}
Network densification combined with universal resources reuse promises significant capacity gains and thus is expected to be a key pillar for future 5G radio access networks. However, such ultra-dense infrastructure deployments lead to challenging interference environments, questioning linear network capacity scaling. To this end, network-wise coordination is considered the key enabler for harnessing the reuse gains. This work is a first attempt to understand the implications of network densification to a future radio access network, provide reference optimization models as well as algorithmic solutions. More specifically, we considered the joint pairing, partitioning and power coordination problem, and we derived for the first time (to the best of our knowledge) an exact solution approach leveraging integer linear programming. We also examined a widely applicable special setting assuming fixed pairing among access and serving nodes, and showed the proposed solution close-to-optimal performance. Due to the complexity of the integer linear programming models, we also developed an algorithmic framework involving two greedy solutions achieving most of the optimal coordination gains with gradually decreasing complexity. We finally applied the optimal and suboptimal solution approaches to various UDN system setups, and demonstrated the significant network capacity enhancements achieved through intelligent coordination, explored rate scaling, and identified optimal partitioning parameterization for the ever-densified UE and infrastructure nodes population.

Based on the insight obtained from the research presented in this paper, several limitations and trade-offs were discovered that should be adequately addressed in order for UDNs to constitute a robust 5G solution for the capacity crunch. First, overhead signaling for estimating and reporting channel quality increases linearly with network densification, and at some point may become prohibitive; to this end, efficient limited feedback techniques are crucial. Secondly, global network coordination assumed in this work could be enabled by a future cloud-based network architecture~\cite{NiCo13}, but, in practice, cluster-based coordination resulting to excess inter-cluster interference may further limit rate scaling~\cite{LoHe13}. Extending the analysis to multi-antenna access nodes and terminals is also critical. As a final remark, the proposed framework explored the coordination capabilities of ANs, assuming exchange of information solely in the signaling plane. Cooperation capabilities (through CoMP or Distributed/Network MIMO) on the other hand, exploiting data-plane information exchange also are highly promising for future UDNs, since they do not only avoid interference but they exploit it to improve performance~\cite{RaKu12}. They come however with severe practical limitations due to backhaul and synchronization requirements that need to be carefully considered~\cite{An13}. Hence, identifying the right mixture of coordination and cooperation techniques given a set of implementation constraints will be key for handling the future challenging interference landscape.

\appendix[Pseudocode of the Power-Aware Partitioning Algorithm
(Sec.\ref{SuboptimalPPP_PowAwarePartitioning})]\label{sec:Appendix1}
\begin{algorithm}
\small
\caption {Power-Aware Suboptimal Partitioning (Section~\ref{SuboptimalPPP_PowAwarePartitioning})}
\begin{algorithmic}[1]
\Statex {/* Initializations:  $\mathcal{S}: $ Pool of unassigned UEs, $\mathcal{L}_n: $ Set of UEs assigned to partition $n$, $\mathcal{M}_n: $  Set of available ANs for partition $n$ */}
\State  $\mathcal{S} = \mathcal{K},\left\langle {{\mathcal{L}_1},{\mathcal{L}_2}, \ldots ,{\mathcal{L}_N}} \right\rangle ,{\mathcal{L}_n} = \emptyset ,\forall n,$

$\left\langle {{\mathcal{M}_1},{\mathcal{M}_2}, \ldots ,{\mathcal{M}_N}} \right\rangle ,{\mathcal{M}_n} = \mathcal{M},\forall n$
\Statex {// Step 1}
\For{$i=1$ to $N$}
\Statex {${\text{// Select triplet}}\left\langle {{n^*},{k^*},{m^*}} \right\rangle {\text{ for current iteration}}$}
\State ${n^*} = i;\left\langle {{k^*}\left( i \right),{m^*}\left( i \right)} \right\rangle  = \operatorname{argmax} \left\{ {{g_{km}}} \right\}_{k \in \mathcal{S}}^{m \in {\mathcal{M}_n}};$
\Statex {// Apply Allocation and Update Sets}
\State {$\mathcal{S} = \mathcal{S}\backslash \left\{ {{k^*}} \right\};{\mathcal{L}_{{n^*}\left( i \right)}} = {\mathcal{L}_{{n^*}\left( i \right)}} \cup \left\{ {{k^*}} \right\};$

${\mathcal{M}_{{n^*}}} = {\mathcal{M}_{{n^*}}}\backslash \left\{ {{m^*}} \right\};$}
\EndFor
\Statex {// Step 2}
\For{$i=N+1$ to $K$}
\State ${\text{Calculate}}\left\langle {{r^*}\left( {{\mathcal{L}_1}} \right),{r^*}\left( {{\mathcal{L}_2}} \right), \ldots ,{r^*}\left( {{\mathcal{L}_N}} \right)} \right\rangle, $ ${\text{           where }}{r^*}\left( {{\mathcal{L}_n}} \right){\text{ is given from Remark~\ref{lem:CommonRateExact} or Remark~\ref{lem:PerronBounds} }}$
\State ${n^*} = \arg \mathop {\max }\limits_n \left\{ {{r^*}\left( {{\mathcal{L}_n}} \right)} \right\};$
\Statex {${\text{// Select triplet}}\left\langle {{n^*},{k^*},{m^*}} \right\rangle {\text{ for current iteration}}$}
\State $\forall k \in \mathcal{S}:\left\langle {m_k^*,{\mathcal{L}_n}^k} \right\rangle  = \left\langle {\arg \mathop {\max }\limits_{m \in {\mathcal{M}_n}} \left\{ {{g_{{k^*}m}}} \right\},{\mathcal{L}_n} \cup \left\{ {{k^*}} \right\}} \right\rangle,$
\State $ {k^*} = \arg \mathop {\max }\limits_{k \in \mathcal{S}} \left\{ {{r^*}\left( {{\mathcal{L}_n}^k} \right)} \right\},{m^*} = m_{{k^*}}^*$
\Statex {Apply Allocation and Update Sets}
\State $\mathcal{S} = \mathcal{S}\backslash \left\{ {{k^*}} \right\};{\mathcal{L}_{{n^*}\left( i \right)}} = {\mathcal{L}_{{n^*}\left( i \right)}} \cup \left\{ {{k^*}} \right\};{\mathcal{M}_{{n^*}}} = {\mathcal{M}_{{n^*}}}\backslash \left\{ {m_k^*} \right\};$
\EndFor
\For{$j=1$ to $N$}
\State $\mathcal{N} = \left\{ j \right\},\mathcal{N}' = \mathcal{N}\backslash \left\{ j \right\}$
\Comment {// Candidate partitions for UE exchange}
\Repeat
\Statex {// Calculate rates per partition before/after swap $\forall \left\langle {n,n'} \right\rangle ,n \in \mathcal{N},n' \in \mathcal{N}',\forall \left\langle {k,k'} \right\rangle ,k \in {\mathcal{L}_n},k' \in {\mathcal{L}_{n'}}$}
\State $r_{current}^* = {r^*}\left( {{\mathcal{L}_n}} \right) + {r^*}\left( {{\mathcal{L}_{n'}}} \right),$

$r_{swap}^* = {r^*}\left( ({{\mathcal{L}_n}\backslash \left\{ k \right\}) \cup \left\{ {k'} \right\}} \right) + {r^*}\left( ({{\mathcal{L}_{n'}}\backslash \left\{ {k'} \right\}) \cup \left\{ k \right\}} \right)$
\Statex {//Check if swap improves rate}
\If{$\Delta r = r_{swap}^* - r_{current}^* > 0$}
\State ${\mathcal{L}_n} = ({\mathcal{L}_n}\backslash \left\{ k \right\}) \cup \left\{ {k'} \right\},{\mathcal{L}_{n'}} = ({\mathcal{L}_{n'}}\backslash \left\{ {k'} \right\}) \cup \left\{ k \right\}$
\Else
\State {Go To next UE pair $\left\langle {k,k'} \right\rangle$}
\EndIf
\Until{all UE pairs have been checked}
\EndFor
\end{algorithmic}
\label{alg:ApproxSol1}
\end{algorithm}
\section*{Acknowledgment}
This work has been performed in the context of the ART-COMP PE7(396)\textit{ ``Advanced Radio Access Techniques for Next Generation Cellular NetwOrks: The Multi-Site Coordination Paradigm"} research project, implemented within the framework ``Supporting Postdoctoral Researchers", and  co-financed by the European Social Fund (ESF) and the Greek State.




\bibliographystyle{IEEEtran}
\bibliography{IEEEfull,gotsis-manuscript-doc}
%

\end{document}